\documentstyle[12pt]{article}

\begin{document}

\baselineskip 22pt

\begin{center}
{\Large \bf Ratios of $B$ and $D$ Meson Decay Constants\\
            in Improved Mock Meson Model}\\
\vspace{1.0cm}
Dae Sung Hwang$^1$ and Gwang-Hee Kim$^2$\\
{\it{Department of Physics, Sejong University, Seoul 143--747,
Korea}}\\
\vspace{2.0cm}
{\bf Abstract}\\
\end{center}

We calculate the ratio $f_{B_s}/f_{B_d}$ by following the
Oakes' method which is based on chiral symmetry breaking,
but we improve his calculation by performing the calculation
of the factor
$<0|{\bar{b}}{\gamma}_5s|B_s>$/$<0|{\bar{b}}{\gamma}_5d|B_d>$
in the mock meson model.
In this calculation we improved the mock meson model by
using the value of the parameter $\beta$ which is obtained
by the variational method in the relativistic quark model.
We also calculate $f_{D_s}/f_{D_d}$, and then
$(f_{B_s}/f_{B_d})\, /\, (f_{D_s}/f_{D_d})$.
In this method we also obtain the ratio $f_{B_s}/f_{D_s}$
and $f_{B_d}/f_{D_d}$ which are important for the
knowledge of CP violation and $B$-$\bar{B}$ mixing.
\\

\vfill

\noindent
$1$: e-mail: dshwang@phy.sejong.ac.kr\\
$2$: e-mail: gkim@phy.sejong.ac.kr
\thispagestyle{empty}
\pagebreak

\baselineskip 22pt

Oakes \cite{oakes}
calculated the ratio $f_{B_s}/f_{B_d}$ based on the usual assumption
that chiral symmetry is broken by quark mass terms
$H_1(x)=\sum m_i{\bar{q}}_i(x)q_i(x)$.
Using the local relation ${\partial}^{\mu}A_{\mu}(x)=-i[Q_5,H_1(x)]$,
he obtained
\begin{equation}
{f_{B_s}\over f_{B_d}}=
\Big( {M_{B_d}\over M_{B_s}}{\Big)}^2
\Big( {m_b+m_s\over m_b+m_d}\Big)
{<0|{\bar{b}}{\gamma}_5s|B_s> \over <0|{\bar{b}}{\gamma}_5d|B_d>},
\label{f1}
\end{equation}
where the decay constants are defined by
\begin{equation}
<0|{\bar{b}}{\gamma}^{\mu}{\gamma}_5s|B_s({\bf{K}})>=if_{B_s}K^{\mu},\ \ \
<0|{\bar{b}}{\gamma}^{\mu}{\gamma}_5d|B_d({\bf{K}})>=if_{B_d}K^{\mu}.
\label{f2}
\end{equation}
Through the symmetry consideration, he took the value of
$<0|{\bar{b}}{\gamma}_5s|B_s>$/$<0|{\bar{b}}{\gamma}_5d|B_d>$
as 1, and then obtained the results
\begin{equation}
f_{B_s}/f_{B_d}=0.989,\ \ \
f_{D_s}/f_{D_d}=0.985,\ \ \
(f_{B_s}/f_{B_d})\, /\, (f_{D_s}/f_{D_d})=1.004.
\label{f3}
\end{equation}

Because of the light quark SU(3) flavor symmetry the factor
$<0|{\bar{b}}{\gamma}_5s|B_s>$/$<0|{\bar{b}}{\gamma}_5d|B_d>$
is almost 1.
However, since we are interested in the ratios in (\ref{f3})
which are by themselves very close to 1,
the deviation of the above factor from 1 is still
important even though it is very small.
Therefore it would be nice if we could calculate the small correction
of the symmetry consideration.
In this Letter we calculate this factor in the mock meson model of
Capstick, Godfrey, and Isgur \cite{gi,godf,cg},
and improve the Oakes' results in (\ref{f3}).
We also apply this approach to the calculations of
$f_{B_s}/f_{D_s}$ and $f_{B_d}/f_{D_d}$,
where the calculations of the above factor are essential
since we can not apply the light quark flavor symmetry
in these cases.
The knowledge of these ratios are important since they
allow us to get the values of $f_{B_s}$ and
$f_{B_d}$ from the experimentally obtained
value of $f_{D_s}$.
The information of $f_{B_s}$ and $f_{B_d}$ is very important,
since it is crucial for the magnitude of CP violation
and the size of $B$-$\bar{B}$ mixing.

The mock meson state is represented by \cite{gi,godf,cg}
\begin{equation}
|M({\bf{K}})>=\int d^3p\, \Phi ({\bf{p}})\,
{\chi}_{s{\bar{s}}}\,
{\phi}_{c{\bar{c}}}\,
|q({m_q\over m}{\bf{K}}+{\bf{p}},s)\,
{\bar{Q}}({m_{\bar{Q}}\over m}{\bf{K}}-{\bf{p}},{\bar{s}})>,
\label{f4}
\end{equation}
where $\bf{K}$ is the mock meson momentum,
$m\equiv m_q+m_{\bar{Q}}$,
and $\Phi ({\bf{p}})$, ${\chi}_{s{\bar{s}}}$, and
${\phi}_{c{\bar{c}}}$ are momentum, spin, and color wave
functions respectively.
We take the momentum wave function $\Phi ({\bf{p}})$ as a
Gaussian wave function
\begin{equation}
\Phi ({\bf{p}})={1\over ( \sqrt{\pi} \beta )^{3/2}}
e^{-{\bf{p}}^2/2{\beta}^2}.
\label{f5}
\end{equation}
We note that we normalized the mock meson state
$|M({\bf{K}})>$ according to
$<M({\bf{K}}')|M({\bf{K}})>={\delta}^3({\bf{K}}'-{\bf{K}})$
\cite{iw},
and we consider only pseudoscalar mesons in the present work.
In this normalization of the mock meson state,
(\ref{f2}) becomes \cite{iw}
\begin{equation}
<0|{\bar{b}}{\gamma}^{\mu}{\gamma}_5s|B_s({\bf{K}})>=
{if_{B_s}K^{\mu}\over {\sqrt{2E_{B_s}}}},\ \ \
<0|{\bar{b}}{\gamma}^{\mu}{\gamma}_5d|B_d({\bf{K}})>=
{if_{B_d}K^{\mu}\over {\sqrt{2E_{B_d}}}}.
\label{f6}
\end{equation}
By applying the local relation
${\partial}^{\mu}A_{\mu}(x)=-i[Q_5,H_1(x)]$
to (\ref{f6}), we get
\begin{equation}
f_{B_s}={{\sqrt{2E_{B_s}}}(m_b+m_s)\over (M_{B_s})^2}
<0|{\bar{b}}{\gamma}_5s|B_s({\bf{K}})>,
\label{f7}
\end{equation}
and a similar expression for $f_{B_d}$.

Then in the rest frames of the mesons we have
\begin{equation}
{f_{B_s}\over f_{B_d}}=
\Big( {M_{B_d}\over M_{B_s}}{\Big)}^{3/2}
\Big( {m_b+m_s\over m_b+m_d}\Big)
{<0|{\bar{b}}{\gamma}_5s|B_s({\bf{0}})>
\over <0|{\bar{b}}{\gamma}_5d|B_d({\bf{0}})>},
\label{f8}
\end{equation}
where
\begin{eqnarray}
<0|{\bar{Q}}{\gamma}_5q|P_{q{\bar{Q}}}({\bf{0}})>
\equiv g(\beta )
&=&2{\sqrt{3}}\int d^3p\,
\Phi ({\bf{p}})\Big(
{E_q+m_q\over 2E_q}\, {E_{\bar{Q}}+m_{\bar{Q}}\over 2E_{\bar{Q}}}
{\Big)}^{1/2}
\nonumber\\
& & \ \ \ \ \ \ \ \ \ \
\times\Big( \, 1\, +\,
{{\bf{p}}^2\over (E_q+m_q)(E_{\bar{Q}}+m_{\bar{Q}})}\, \Big) .
\label{f9}
\end{eqnarray}
We note that the power of $M_{B_d} / M_{B_s}$ in (\ref{f8})
is 3/2 instead of 2 in (\ref{f1}), since we use the
normalization
$<M({\bf{K}}')|M({\bf{K}})>={\delta}^3({\bf{K}}'-{\bf{K}})$,
whereas Oakes used the more common normalization
$<M({\bf{K}}')|M({\bf{K}})>=2E\, {\delta}^3({\bf{K}}'-{\bf{K}})$.
We also note that the sign inside the last parenthesis
in (\ref{f9}) is plus, whereas that in Eq. (3) of Ref. \cite{godf}
is minus.
This difference originates from the fact that we consider
the pseudoscalar quantity of the left hand side of (\ref{f9}),
whereas Ref. \cite{godf} considered the zeroth component
of the four vector
$<0|{\bar{q}}{\gamma}^{\mu}(1-{\gamma}_5)|M({\bf{K}})>$.
In order to calculate the ratio of the decay constants
from (\ref{f8}),
we should calculate the matrix element
$<0|{\bar{Q}}{\gamma}_5q|P_{q{\bar{Q}}}({\bf{0}})>$.
This matrix element is a function of $\beta$, which we call
$g(\beta )$.
We performed numerical calculations for
the $g(\beta )$ of $B_s$, $B_d$, $D_s$, and $D_d$ respectively,
and present the results in Fig. 1.

Then the problem is what values we should use for $\beta$.
Capstick and Godfrey used the values of $\beta$ obtained from
the effective harmonic oscillator potential \cite{godf,cg}.
Here, we improve this treatment by applying the variational
method to the relativistic hamiltonian \cite{hkn,hk}
\begin{equation}
H={\sqrt{{\bf p}^2+{m_Q}^2}}+{\sqrt{{\bf p}^2+{m_q}^2}}+V(r),
\label{b1}
\end{equation}
where ${\bf r}$ and ${\bf p}$ are the relative coordinate and its
conjugate momentum.
The hamiltonian in (\ref{b1}) represents the energy of the meson in
the meson rest frame, since in this reference frame the momentum
of each quark is the same in magnitude as
that of the conjugate momentum of the relative coordinate.
In Ref. \cite{hk} we obtained the values of $\beta$
which minimize the expectation value of the hamiltonian in (\ref{b1}).
We considered
six different potentials for $V(r)$ in (\ref{b1}),
and the averages of the minimizing $\beta$ values, $\bar{\beta}$,
obtained from six potentials
are given by \cite{hk}
\begin{equation}
{\bar{\beta}}_{B_s}=0.579\ {\rm{GeV}},\ \ \
{\bar{\beta}}_{B_d}=0.558\ {\rm{GeV}},\ \ \
{\bar{\beta}}_{D_s}=0.492\ {\rm{GeV}},\ \ \
{\bar{\beta}}_{D_d}=0.475\ {\rm{GeV}}.
\label{f11}
\end{equation}
In the above calculations we used the current quark masses of the
light quarks as $m_d$=9.9 MeV and $m_s$=199 MeV,
which were given by Dominguez and Rafael \cite{dr}.

By using the values in (\ref{f11}) for $\bar{\beta}$,
we can obtain from the $g(\beta )$ in Fig. 1 the values of
the matrix elements
$g({\bar{\beta}} )\equiv <0|{\bar{Q}}{\gamma}_5q|P_{q{\bar{Q}}}({\bf{0}})>$
in (\ref{f9}),
which are given by
\begin{equation}
g_{B_s}(0.579)= 8.657,\ \ \
g_{B_d}(0.558)= 7.494,\ \ \
g_{D_s}(0.492)= 7.418,\ \ \
g_{D_d}(0.475)= 6.533.
\label{f12}
\end{equation}
Then, using the meson \cite{rpp}
and the current quark \cite{dr} masses given by
\begin{eqnarray}
& &M_{B_s}= 5.375\, {\rm{GeV}},\
M_{B_d}= 5.279\, {\rm{GeV}},\
M_{D_s}= 1.969\, {\rm{GeV}},\
M_{D_d}= 1.869\, {\rm{GeV}},
\nonumber\\
& &m_b= 4.397\, {\rm{GeV}},\
m_c= 1.306\, {\rm{GeV}},\
m_s= 0.199\, {\rm{GeV}},\
m_d= 0.0099\, {\rm{GeV}},
\label{f13}
\end{eqnarray}
we obtain the following ratios of the decay constants
from (\ref{f8}).
\begin{equation}
{f_{B_s}\over f_{B_d}}=1.173,\ \ \
{f_{D_s}\over f_{D_d}}=1.201,\ \ \
{f_{D_s}\over f_{B_s}}=1.266,\ \ \
{f_{D_d}\over f_{B_d}}=1.236.
\label{f14}
\end{equation}
When we use the values of the ratios in (\ref{f14})
together with the experimental value of $f_{D_s}$
given by $f_{D_s}=265\pm 68$ MeV
\cite{hk,rpp,wa75,cleo94},
we obtain $f_{D_d}=221\pm 57$ MeV,
$f_{B_s}=209\pm 54$ MeV, and $f_{B_d}=178\pm 46$ MeV.

In order to compare our results in (\ref{f14}) with the
Oakes' original approach, we calculated the ratios
using the formula (\ref{f1}) with our values of the meson
and current quark masses in (\ref{f13}).
Here, we took the last factor in (\ref{f1}) as 1, as
Oakes did.
The results are presented in the row 2 of Table 1.
As we see in Table 1, our results of
the ratios $f_{B_s}/f_{B_d}$ and $f_{D_s}/f_{D_d}$
have been enhanced significantly compared with the
Oakes' results, since we have taken care of
the meson structure through (\ref{f8}) and (\ref{f9}).
However, the double ratio
$(f_{B_s}/f_{B_d})/(f_{D_s}/f_{D_d})$
remains the same.

For comparisons, we summarize the results
of various different calculations in Table 1.
Grinstein did the calculation using the heavy quark
effective theory, and Dominguez using the QCD sum rules.
The row 6 is the results of the original calculations
of Capstick and Godfrey.
They used the values of the parameter $\beta$ given by
the effective harmonic oscillator potential,
whereas we used the values given by the variational
method in the relativistic quark model.
Another important difference between their and our
calculations is that they used the formula (\ref{f2})
directly and had the ambiguity of which component
of the four vector should be used to
obtain $f_P$ as they discussed \cite{cg},
whereas we followed the Oakes' approach based on chiral
symmetry breaking and then arrived at the formula
(\ref{f8}) which has no such ambiguity.
In the row 7 we present the results which the authors
of this Letter obtained in the relativistic quark
model \cite{hk}.
The rows 8-11 are the
results of the lattice calculations, where
the third column was estimated from their announced
values of the first
and second columns, and the fourth and fifth columns were
estimated from their results of
$f_{D_s}$, $f_{B_s}$, $f_{D_d}$, and $f_{B_d}$.\\

\pagebreak

\noindent
{\em Acknowledgements} \\
\indent
This work was supported
in part by the Basic Science Research Institute Program,
Ministry of Education, Project No. BSRI-94-2414,
and in part by Daeyang Foundation at Sejong University in 1994.\\

\pagebreak

\pagebreak

\vspace*{3.5cm}

\begin{table}[h]
\vspace*{0.7cm}
\hspace*{-2.5cm}
\begin{tabular}{|c|c|c|c|c|c|c|}   \hline
    &     &$f_{B_s}/f_{B_d}$&$f_{D_s}/f_{D_d}$&
${(f_{B_s}/f_{B_d})\over (f_{D_s}/f_{D_d})}$&
$f_{D_s}/f_{B_s}$&$f_{D_d}/f_{B_d}$\\  \hline
1&This Work&1.173&1.201&0.977&1.266&1.236\\

2&Oakes Modif.&1.006&1.030&0.977& --- & --- \\

3&Oakes \cite{oakes}&0.989&0.985&1.004& --- & --- \\

4&Grinstein \cite{grin}& --- & --- &0.967& --- & --- \\

5&Dominguez \cite{doming}&1.22$\pm .02$&1.21$\pm .06$
&1.01$\pm .05$& ---  & --- \\

6&Cap. Godf. \cite{cg} &1.35$\pm .18$&1.21$\pm .13$
&1.12$\pm .19$&1.38$\pm .16$&1.55$\pm .20$\\

7&Hwang Kim \cite{hk}&1.053&1.045&1.008&1.251&1.261\\

8& ELC  \cite{ELC} &1.08$\pm .06$&1.08$\pm .02$&1.00$\pm .06$
&1.03$\pm .22$&1.02$\pm .21$\\

9&UKQCD\cite{UKQCD}&1.22${\, }^{+.04}_{-.03}$&1.18$\pm .02$
&1.03${\, }^{+.04}_{-.03}$&1.09${\, }^{+.04+.42}_{-.03-.06}$
&1.16${\, }^{+.05+.46}_{-.05-.14}$\\

10& BLS  \cite{BLS} &1.11$\pm .02\pm .05$&1.11$\pm .02\pm .05$
&1.00$\pm .03\pm .06$&1.11$\pm .06\pm .27$&1.11$\pm .08\pm .30$\\

11& MILC \cite{MILC}&1.13(2)(9)(4)&1.09(1)(4)(4)&1.04(2)(9)(5)
&1.18(3)(17)(13)&1.22(5)(17)(19)\\
                    \hline
\end{tabular}
\caption{The ratios of the decay constants from
different calculations.}
\end{table}

\pagebreak

\vspace*{3.5cm}
\vspace*{14.5cm}
\noindent
Fig. 1.
$g_{B_s}(\beta )\equiv <0|{\bar{b}}{\gamma}_5s|B_s({\bf{0}})>$,
$g_{B_d}(\beta )\equiv <0|{\bar{b}}{\gamma}_5d|B_d({\bf{0}})>$,
$g_{D_s}(\beta )\equiv <0|{\bar{c}}{\gamma}_5s|D_s({\bf{0}})>$, and
$g_{D_d}(\beta )\equiv <0|{\bar{c}}{\gamma}_5d|D_d({\bf{0}})>$,
as functions of the parameter $\beta$.

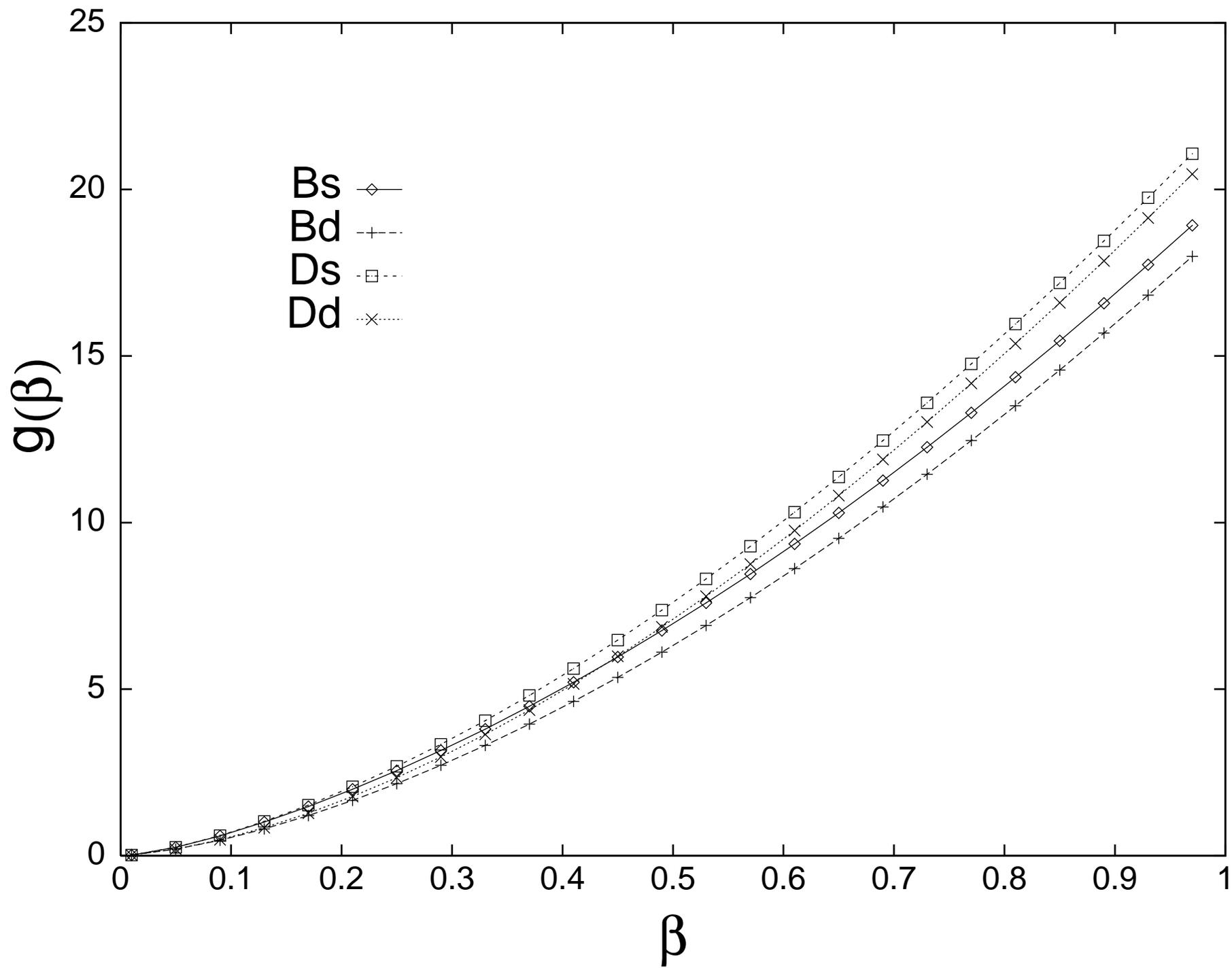
\end{document}